\documentclass{elsart3p}

\usepackage{graphicx}

\usepackage{amssymb}

\begin{document}

\begin{frontmatter}



\title{Temperature and Humidity Dependence of Air Fluorescence Yield measured by AIRFLY}



\author[Chi]{{\bf AIRFLY Collaboration}: M. Ave},
\author[CZ1]{ M. Bohacova},
\author[LNF]{ B. Buonomo},
\author[Chi]{ N. Busca},
\author[Chi]{ L. Cazon},
\author[ANL]{ S.D. Chemerisov},
\author[ANL]{ M.E. Conde},
\author[ANL]{ R.A. Crowell},
\author[AqU]{ P. Di Carlo},
\author[RomeU]{ C. Di Giulio},
\author[CZ2]{ M. Doubrava},
\author[LNF]{ A. Esposito},
\author[Sant]{ P. Facal},
\author[ANL]{ F.J. Franchini},
\author[KarlsruheU]{ J. H\"orandel\thanksref{add}},
\author[CZ1]{ M. Hrabovsky},
\author[AqU]{ M. Iarlori},
\author[ANL]{ T.E. Kasprzyk},
\author[KarlsruheU]{ B. Keilhauer},
\author[FZK1]{ H. Klages},
\author[FZK2]{ M. Kleifges},
\author[ANL]{ S. Kuhlmann},
\author[LNF]{ G. Mazzitelli},
\author[CZ1]{ L. Nozka},
\author[KarlsruheU]{ A. Obermeier},
\author[CZ1]{ M. Palatka},
\author[AqU]{ S. Petrera},
\author[RomeU]{ P. Privitera\corauthref{cor1}},
\ead{priviter@roma2.infn.it}
\author[CZ1]{ J. Ridky},
\author[AqU]{ V. Rizi},
\author[RomeU]{ G. Rodriguez},
\author[AqU]{ F. Salamida},
\author[CZ1]{ P. Schovanek},
\author[ANL]{ H. Spinka},
\author[RomeU]{ E. Strazzeri},
\author[Mun]{ A. Ulrich},
\author[ANL]{ Z.M. Yusof},
\author[CZ2]{ V. Vacek},
\author[RomeS]{ P. Valente},
\author[RomeU]{ V. Verzi},
\author[FZK1]{ T. Waldenmaier}
\footnotesize
\address[Chi]{ University of Chicago, Enrico Fermi Institute, 5640 S. Ellis Ave., Chicago, IL 60637, United States}
\address[CZ1]{Institute of Physics of the Academy of Sciences of the
Czech Republic, Na Slovance 2, CZ-182 21 Praha 8, Czech
Republic}
\address[LNF]{Laboratori Nazionali di Frascati dell'INFN, INFN, Sezione di Frascati, Via Enrico Fermi 40, Frascati, Rome 00044, Italy }
\address[ANL]{Argonne National Laboratory, Argonne, IL 60439 United States}
\address[AqU]{Dipartimento di Fisica dell'Universit\`{a} de l'Aquila and
INFN, Via Vetoio, I-67010 Coppito, Aquila, Italy}
\address[RomeU]{Dipartimento di Fisica dell'Universit\`{a} di Roma Tor
Vergata and Sezione INFN, Via della Ricerca Scientifica, I-00133 Roma, Italy}
\address[CZ2]{Czech Technical University, Technicka 4, 16607 Praha 6, Czech Republik}
\address[Sant]{ Departamento de F\'{\i}sica de Part\'{\i}culas, Campus Sur, Universidad, E-15782 Santiago de Compostela, Spain}
\address[KarlsruheU]{ Universit\"{a}t Karlsruhe (TH), Institut f\"{u}r Experimentelle Kernphysik (IEKP), Postfach 6980, D - 76128 Karlsruhe, Germany }
\address[FZK1]{Forschungszentrum Karlsruhe, Institut f\"{u}r Kernphysik,
Postfach 3640, D - 76021 Karlsruhe, Germany}
\address[FZK2]{Forschungszentrum Karlsruhe, Institut f\"{u}r Prozessdatenverarbeitung und Elektronik, Postfach 3640, D - 76021 Karlsruhe, Germany}
\address[Mun]{Physik Department E12, Technische Universit\"{a}t Muenchen,
James Franck Str. 1, D-85748 Garching, Germany}
\address[RomeS]{Sezione INFN di Roma 1, Ple. A. Moro 2, I-00185 Roma, Italy}
\corauth[cor1]{corresponding author}
\thanks[add]{now at Department of Astrophysics, Radboud University Nijmegen, Nijmegen, The Netherlands}

\begin{abstract}
The fluorescence detection of ultra high energy cosmic rays requires a detailed knowledge of the fluorescence light emission from nitrogen molecules over a wide range of atmospheric parameters, corresponding to altitudes typical of the cosmic ray shower development in the atmosphere. We have studied the temperature and humidity  dependence of the fluorescence light spectrum excited by MeV electrons in air. Results for the 313.6 nm, 337.1 nm, 353.7 nm and 391.4 nm bands are reported in this paper. We found that the temperature and humidity dependence of the quenching process changes the fluorescence yield  by a sizeable amount (up to 20\%) and its effect must be included for a precise estimation of the energy of ultra high energy cosmic rays.
\end{abstract}

\journal{5th Fluorescence Workshop, Madrid, 2007}
\end{frontmatter}

\section{Introduction}
\label{intro}
     The detection of ultra high energy ($\gtrapprox 10^{18}$ eV) cosmic rays using nitrogen fluorescence light emission from extensive air showers (EAS) is a well established technique, used by the Fly's Eye \cite{flyseye}, HiRes \cite{hires}, and
Pierre Auger Observatory \cite{auger} experiments, and for the
Telescope Array \cite{telaray} under construction.  It
has also been proposed for the satellite-based EUSO and OWL projects. 
Excitation of atmospheric nitrogen by EAS charged particles induces fluorescence emission, mostly in the wavelength range between 300 to 430 nm. 
Information on the longitudinal EAS development
can be obtained by fluorescence telescopes by recording the light
intensity as a function of time and incoming direction. However, the fluorescence light yield from EAS charged particles must be
well known at each point within the shower, and corrections applied for
atmospheric effects between the shower and the telescope for an
accurate primary energy determination. Thus, the intensities of the fluorescence bands should be measured over a range of air pressure and temperature corresponding to altitudes up to about 16 km, the typical elevation of EAS development in the atmosphere. The presence of humidity will also affect the fluorescence yield, the effect being more important for satellite experiments which will detect showers over the oceans.

The AIRFLY (AIR FLuorescence Yield) collaboration is pursuing an extensive measurement program of the fluorescence light yield with significantly improved precision with respect to previous experiments \cite{flbib}.
A precise measurement of the pressure dependence of the air fluorescence band intensities has been published \cite{airflyAP}, and is summarized in \cite{obermeier}. Measurements of the fluorescence yield dependence on the electron kinetic energy from keV to GeV are presented in \cite{maximo}. The status of AIRFLY measurement of the absolute fluorscence yield is presented in \cite{absolute}. The data reported here address the measurement of the fluorescence yield spectrum dependence on air temperature and humidity.

The fluorescence yield of a band of wavelength $\lambda$ at a given pressure $p$ and temperature $T$ can be written as \cite{airflyAP}:
\begin{eqnarray}
 Y_{air}(\lambda,p,T)&=& Y_{air}(337,p_0,T_0) \nonumber \cdot  I_{\lambda}(p_0,T_0)  \nonumber \\ && \cdot \frac{1+\frac{p_0}{p'_{air}(\lambda,T_0)}}{1+\frac{p}{p'_{air}(\lambda,T_0)\sqrt{\frac{T}{T_0}}\frac{H_\lambda(T_0)}{H_\lambda(T)}}} ,
\label{eq:phish}
\end{eqnarray}
 where  $Y_{air}(337,p_0,T_0)$ is the absolute yield of the 337 nm band at pressure $p_0$ and temperature $T_0$ (in photons emitted per MeV of energy deposited), $I_{\lambda}(p_0,T_0)$ is the $\lambda$ band intensity relative to the 337 nm band and $p'_{air}(\lambda,T_0)$  is the band quenching reference pressure. $H_\lambda(T)$ has been introduced to take into account a possible temperature dependence of the collisional cross sections.

The effect of humidity in the fluorescence yield can be introduced by substituting in Eq. (\ref{eq:phish}):
\begin{eqnarray}
\frac{1}{p'_{air}} &\rightarrow& \frac{1}{p'_{air}} \left( 1-\frac{p_{h}}{p} \right)  + \frac{1}{p'_{\rm{H_2O}}}\frac{p_{h}}{p},
\label{eq:hum}
\end{eqnarray}
where $p_{h}$ is the water vapour partial pressure and $p'_{\rm{H_2O}}$ is the water vapour collisional quenching pressure.

The experimental method used for the measurement of the temperature and humidity dependence is described in Section \ref{sec:expmethod}. The results obtained for the humidity and temperature dependence of four major fluorescence bands (313.6 nm, 337.1 nm, 353.7 nm and 391.4 nm) are presented in Sections \ref{sec:temperature} and \ref{sec:humidity}, respectively. Implications for the energy determination of ultra high energy cosmic rays of the AIRFLY measurements reported in this paper are discussed in Section \ref{sec:implications}. Conclusions are given in Section \ref{sec:conclusions}.

\section{Experimental method}
\label{sec:expmethod}

     Measurements of both the humidity and temperature dependence of 
the fluorescence light yield were performed at the Argonne National 
Laboratory Chemistry Division's Van de Graaff (VdG) electron 
accelerator \cite{VdG} \cite{VdGsor}.  It was operated in the DC 
current mode with typical beam currents of approximately 10 $\mu A$ 
and nominal beam kinetic energy of 3.0 MeV.  After a $30^{\circ}$ 
bend in an electromagnet, the beam was focused near the exit from 
the accelerator vacuum, which consisted of a 35 mm diameter and 
0.152 mm thick dura-aluminum window.  For beam tuning, the beam 
spot of about 6 mm diameter was viewed with a quartz plate and 
mirror close to the exit window, and a camera.  The quartz plate and 
mirror were removed during data collection. For each humidity or temperature measurement, spectra from 284 - 429 nm were recorded with a spectrograph. The beam current was also measured by a Faraday cup located in the accelerator vacuum.

\subsection{Humidity Measurements}
\label{sec:humsec}

     The humidity measurements occurred in the same setup shown in 
Fig. 1 of Ref.~\cite{airflyAP}.  The electron beam entered the 
pressure chamber through a 0.50 mm thick beryllium window, 
traversed 378 mm of gas, and left through a 0.1 mm thick aluminum 
window.  Fluorescence light produced in the gas left the pressure 
chamber through a quartz window; it was collected by a 10 m long, 
1.5 mm diameter pure silica core optical fiber, which brought the 
light to a spectrograph. The aluminum spherical mirror in 
\cite{airflyAP} was not used for these measurements.  The 
spectrograph was located behind a concrete block wall with 
additional lead shielding to protect it from radiation produced 
by the VdG.

     A remotely-controlled vacuum and gas handling system was used 
with the pressure chamber.  A dry scroll vacuum pump \cite{pump}
was used, and the chamber pressure \cite{gage} and relative 
humidity (rH) (Sensorika probe HTP-9511533 \cite{humsen}) were measured at the pump-out port.  A high 
purity dry gas (78.0\% nitrogen, 21.0\% oxygen, 1.0\% argon) passed 
through a bubbler containing high purity water before entering the 
pressure chamber.

     Air fluorescence spectra were recorded by an Oriel MS257$^{TM}$
spectrograph \cite{spec}, described in detail in \cite{airflyAP}.  
Calibrations with a mercury pencil lamp (Oriel no. 60635) and a 
quartz tungsten halogen lamp (Oriel no. 63350) were performed.  
In a fluorescence run the spectrograph collected data first in 
the wavelength range 284-369 nm and then in the range 344-429 nm.  
For each range, 50 spectra of 1 second integration time were taken 
in an automated sequence.  The beam current from the Faraday cup 
was recorded before and after each sequence of spectra collected.

\subsection{Temperature Measurements}
\label{sec:tempsec}

\begin{figure}
\begin{center}
\includegraphics [width=0.48\textwidth]{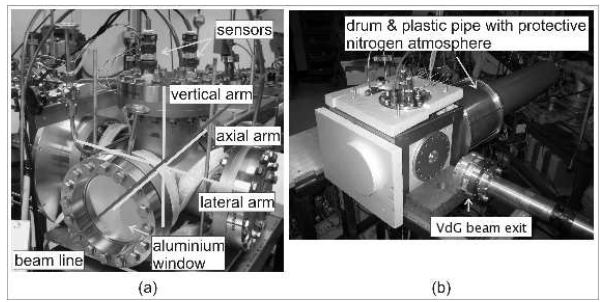}
\end{center}
\caption{The temperature chamber at the Van de Graaff: a) the chamber on the beam line as it appear before mounting the polysterene box, b) the chamber inside the polystyrene box, with the protective drum and pipe also in place.}
\label{fig:VdG}
\end{figure}

     A special temperature chamber was constructed for the temperature 
measurements; see Fig. \ref{fig:VdG}.  It consisted of a stainless 
steel pipe with two horizontal arms and one vertical arm.  The electron 
beam entered and exited the chamber through 100 $\mu m$ thick 
aluminum windows.  One horizontal arm had a stainless steel plate 
blankoff, and the other had the same quartz window as the pressure 
chamber.  The gas length traversed by the electron beam in the 
temperature chamber was approximately 220 mm.

     The vertical arm of the temperature chamber contained feed-thrus 
for five sensors: two relative humidity (moisture) sensors (Sensorika 
HS 2Ta \cite{humsen}), a pressure sensor (Sensortechnics BT/PTU7000 
\cite{pressen}), and two temperature sensors (Sensorika Pt 10000 
\cite{humsen}).  The humidity and pressure sensors were located near 
the vacuum flange with the feed-thrus, about 120 mm above the beam 
line.  One temperature sensor was near beam height, but offset 
perpendicular to the beam and away from the quartz window about 3 cm, 
and the other was approximately 70 mm above beam height.  Signals 
from all sensors were collected in a portable data acquisition system 
 including a calibrated signal converter, and then sent to a 
PC to monitor the chamber conditions.

     The temperature chamber body was enclosed in a polystyrene box 
with aluminum supporting structure.  The box was filled with dry ice 
($CO_2$) for cooling of the chamber down to -40 $^{\circ}$C.  A 
2 m long strip heater (48 V, 325 W) wrapped around the chamber 
body was used to regulate the temperature inside.  

     In order to avoid buildup of frost on the quartz window, a 
protective drum and pipe were added to the temperature chamber.  
These covered the window and end of the optical fiber.  Boil-off 
gas from a liquid nitrogen dewar was continuously fed through the 
protective drum and pipe system.

     The same remotely-controlled vacuum and gas handling system, 
silica core optical fiber, Faraday cup, and spectrograph were used 
as for the humidity measurements.

\section{Temperature dependence}
\label{sec:temperature} 

Fluorescence yield measurements were performed in the temperature range between 240~K and 310~K in dry air. After allowing for temperature to stabilize in the chamber, a fluorescence spectrum was taken. The fluorescence band intensities were estimated by integrating the CCD counts in a wavelength interval around each band. A detailed account of the procedure can be found in \cite{airflyAP}. 
 For each fluorescence band $\lambda$, the corresponding fluorescence signal $S_{air}(\lambda)$ was obtained from the ratio of the number of counts in the band interval to the beam current. 

\begin{figure}
\begin{center}
\includegraphics [width=0.48\textwidth]{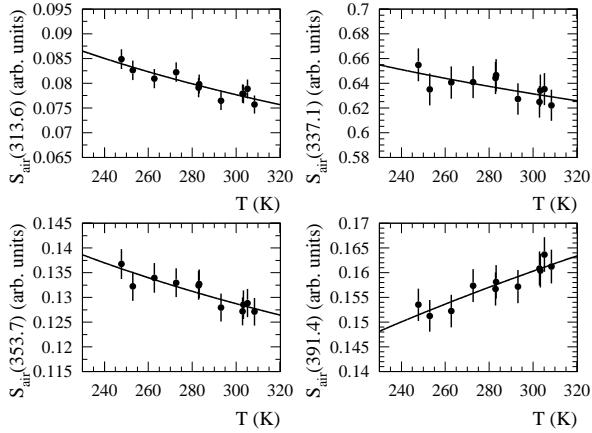}
\end{center}
\caption{Temperature dependence of the air fluorescence signal for the 313.6 nm, 337.1 nm, 353.7 nm and 391.4 nm bands}\label{fig:temp}
\end{figure}

The temperature dependence of the fluorescence signal of the 313.6 nm, 337.1 nm, 353.7 nm and 391.4 nm bands is shown in Fig. \ref{fig:temp}, together with the result of a fit to the data with the following ansantz in Eq. (\ref{eq:phish}):
\begin{eqnarray}
\frac{H_\lambda(T)}{H_\lambda(T_0)} = \left( \frac{T}{T_0} \right)^{\alpha_\lambda} .
\label{eq:tdep}
\end{eqnarray}
The fitted values of $\alpha_\lambda$ are reported in Table \ref{tab:table}. Notice that $\alpha_\lambda=0$ indicates that collisional cross sections have no temperature dependence, which has been so far assumed in the application of fluorescence yield measurements to fluorescence detection of ultra high energy cosmic rays. The AIRFLY measurements show a sizable deviation of the temperature dependence of the fluorescence yield from this usual assumption. Also, the value of $\alpha_\lambda$ appears to depend on the wavelength band, with the 391.4 nm band being significantly different from the others. 

\section{Humidity dependence}
\label{sec:humidity} 

\begin{figure}
\begin{center}
\includegraphics [width=0.48\textwidth]{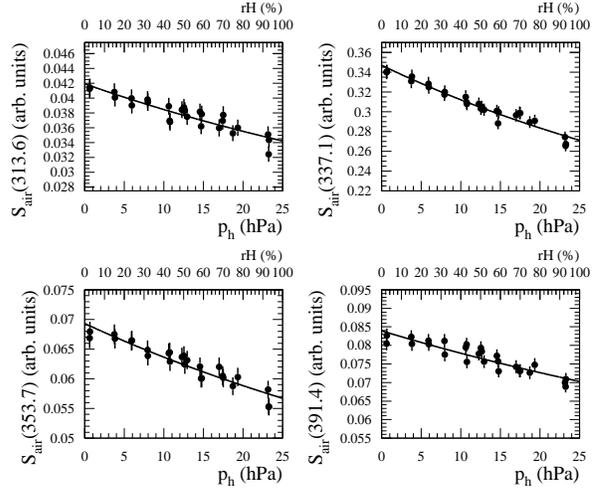}
\end{center}
\caption{Air fluorescence signal as a function of water vapour partial pressure (relative humidity on the top axis) for the 313.6 nm, 337.1 nm, 353.7 nm and 391.4 nm bands.}\label{fig:humid}
\end{figure}

Fluorescence yield measurements were performed for relative humidity ranging from 0 to almost 100\% corresponding to water vapour partial pressures  $p_h$ up to about 25~hPa. After allowing for humidity to stabilize in the chamber, a fluorescence spectrum was taken. For each fluorescence band $\lambda$ , the corresponding fluorescence signal $S_{air}(\lambda)$ was obtained from the ratio of the number of counts in the band interval to the beam current.  The beam current was measured several times during the run with a Faraday cup placed inside the beam pipe.    
The measured fluorescence signal as a function of the water vapour partial pressure $p_h$  for the 313.6 nm, 337.1 nm, 353.7 nm and 391.4 nm bands is shown in Fig. \ref{fig:humid}, together with the result of a fit to the data using Eq. (\ref{eq:hum}).
The fitted values of the water vapour collisional quenching pressure $p'_{\rm{H_2O}}$ are reported in Table \ref{tab:table}. Notice that the quenching effect of water vapour is not negligible, resulting in about 20\% decrease of the fluorescence yield for rH=100\%. 
 \begin{table}[hp]
\centering
\begin{footnotesize}
\begin{tabular}{|c|c|c|}
\hline
\hline
$\lambda$ (nm)  & $\alpha_\lambda$ &  $p'_{\rm{H_2O}}$ (hPa) \\
\hline
   313.6  &$-0.09 \pm 0.10$ &$1.21 \pm 0.13$\\
   337.1  &$-0.36 \pm 0.08$ &$1.28 \pm 0.08$\\
   353.7  &$-0.21 \pm 0.09$&$1.27 \pm 0.12$\\
   391.4  & $-0.80 \pm 0.09$&$0.33 \pm 0.03$\\
\hline
\end{tabular}
\end{footnotesize}
\vskip 0.1truein
\caption{Summary of measured temperature and humidity dependence parameters for a selected group of air fluorescence bands.}
\label{tab:table}.
\end{table}

\section{Implications for the fluorescence detection of ultra high energy cosmic rays}
\label{sec:implications} 
\begin{figure}
\begin{center}
\includegraphics [width=0.48\textwidth]{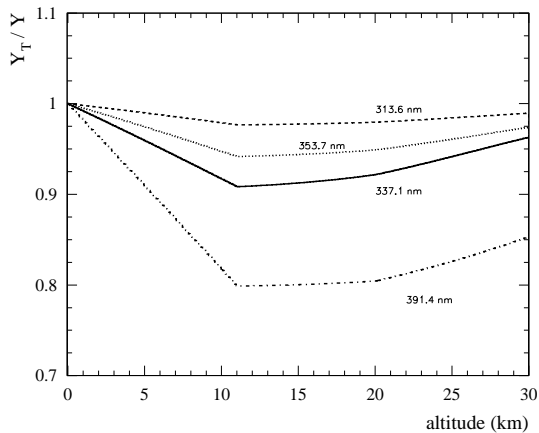}
\end{center}
\caption{Ratio of fluorescence yield with measured $\alpha_\lambda$ to the one with  $\alpha_\lambda=0$: dashed line 313.6 nm; full line 337.1 nm; dotted line 353.7 nm; dashed-dotted 391.4 nm}\label{fig:yieldtemp}
\end{figure}

Experiments employing the fluorescence technique for the detection of ultra high energy cosmic rays  have so far assumed $\alpha_\lambda= p_{h} = 0$ in the treatment of fluorescence yield. The AIRFLY results presented in Sections \ref{sec:temperature} and \ref{sec:humidity} do not support this assumption. In fact, neglecting temperature and humidity effects on the fluorescence yield introduces a sizeable bias in the energy estimation of EAS. In Fig. \ref{fig:yieldtemp}, the ratio of the fluorescence yield with $\alpha_\lambda$ from Table \ref{tab:table} to the one with  $\alpha_\lambda=0$ is shown as a function of altitude. All yields were normalized to have the same value at ground level, and the U.S. 1976 Standard Atmosphere \cite{stda} was used to calculate the pressure and temperature at a given altitude. 
It appears  from  Fig. \ref{fig:yieldtemp} that neglecting the temperature dependence results in an overestimation of the fluorescence yield by an amount going up to $\approx 20\%$ for the 391.4 nm band.

\begin{figure}
\begin{center}
\includegraphics [width=0.48\textwidth]{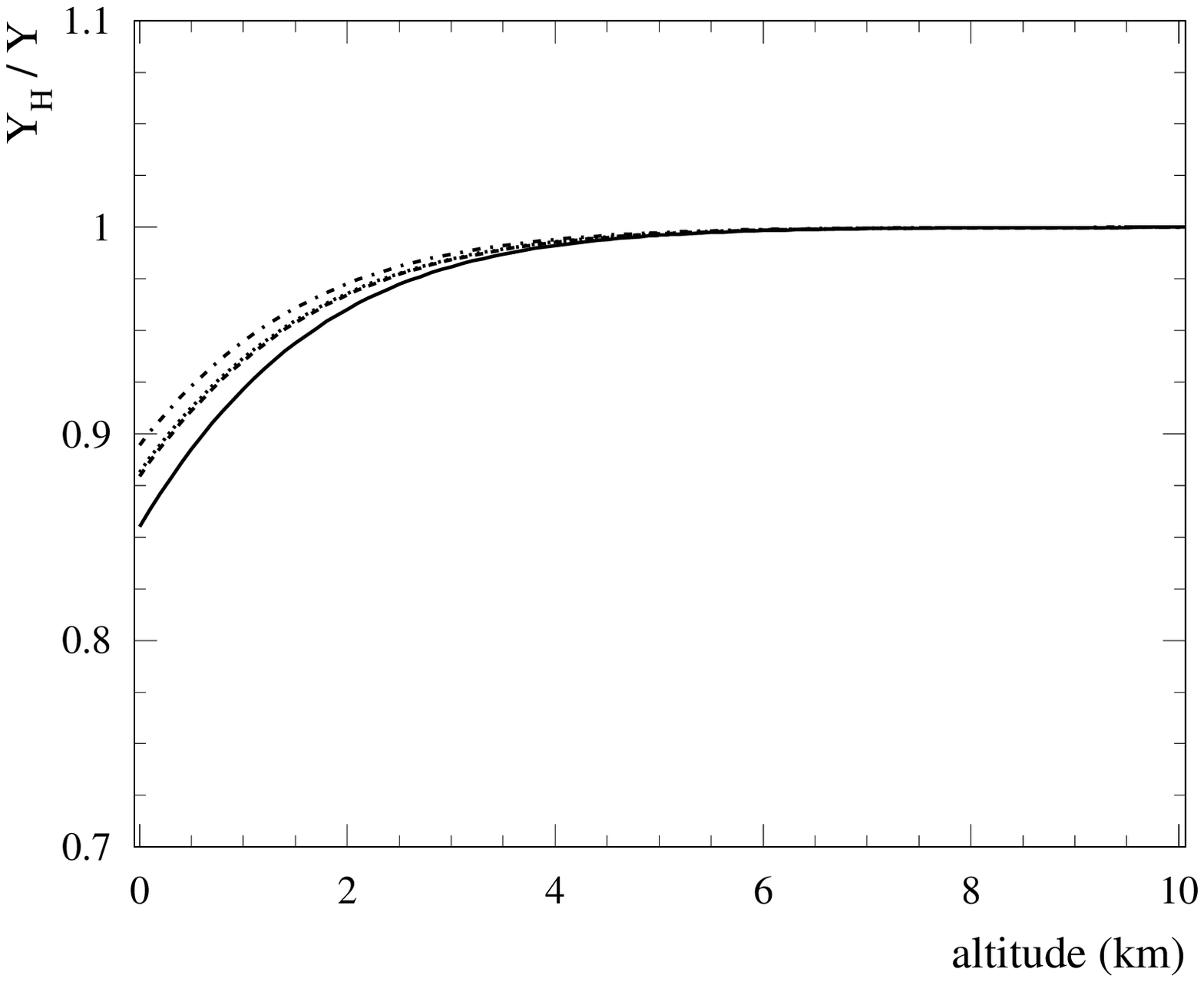}
\end{center}
\caption{Ratio of fluorescence yield with measured $p'_{\rm{H_2O}}$ to the one with $p_{h}=0$: dashed line 313.6 nm; full line 337.1 nm; dotted line 353.7 nm; dashed-dotted 391.4 nm}\label{fig:yieldhum}
\end{figure}

We also investigated the bias introduced by neglecting the humidity effect on the fluorescence yield.  We expect the effect to be more important for satellite experiments which will detect showers over the oceans. In Fig. \ref{fig:yieldhum}, the ratio of the fluorescence yield with $p'_{\rm{H_2O}}$ from Table \ref{tab:table} to the one with  $p_{h}=0$ is shown as a function of altitude, assuming a humidity profile over oceans as given in \cite{ocean}. 
The effect of humidity is sizeable up to about 5 km altitude. 
 
A full estimate of the effect of neglecting temperature and humidity in the fluorescence yield on the EAS energy measurement would need a detailed study, including the EAS energy deposit as a function of altitude, the wavelength dependent attenuation of the emitted fluorescence light in the atmosphere and the sensitivity of the fluorescence detectors. This goes beyond the scope of this paper.
On the other hand, the experimental accuracy on the energy estimation of EAS with the fluorescence detection  technique asks for a systematic uncertainty on the knowledge of the fluorescence yield better than 10\%. Thus, biases of the order of 10\%, as seen in  Figs. \ref{fig:yieldtemp} and \ref{fig:yieldhum}, should not be neglected. 

\section{Conclusions}
\label{sec:conclusions} 
The measurement of the EAS energy with the fluorescence detection technique requires a detailed knowledge of the air fluorescence emission as a function of pressure, temperature and humidity.
The AIRFLY experiment has performed precise measurements of the fluorescence light spectrum excited by MeV electrons in dry air. The relative intensities of 34 fluorescence bands as well as their pressure dependence have been fully reported \cite{airflyAP}, and are summarized in a contribution to this workshop \cite{obermeier}. In this paper, we reported measurements of the temperature and humidity dependence of the 313.6 nm, 337.1 nm, 353.7 nm and 391.4 nm fluorescence bands in air. The temperature dependent measurements of the fluorescence yield, performed in the range between 240~K and 310~K, are the first reported in the literature. Our data show that collisional cross sections are temperature dependent, contrary to the assumption so far made in the UHECR community. 
We also measured the quenching effect of water vapour, resulting in about 20\% decrease of the fluorescence yield for rH=100\%.
The effect of the temperature and humidity dependence of the fluorescence yield on the energy estimation of EAS was studied, and we found that it should not be neglected.  
A full analysis of the AIRFLY data, with the measurement of the temperature and humidity dependence of 34 fluorescence bands, is in preparation.  

\section{Acknowledgments}
We thank the staff of Argonne National Laboratory for their support. This work was also supported by the grant of MSMT CR LC 527 and 1M06002 and ASCR grants AV0Z10100502 and AV0Z10100522. A.\ Obermeier and J.\ R.\ H\"orandel acknowledge the support of VIHKOS, which made the participation at the measurement campaigns possible.









\bibliographystyle{elsart-num}
\bibliography{biblio}

\end{document}